\documentclass[amsmath,aps,showpacs,a4paper,10pt]{revtex4}

 \usepackage{epsf}
 \usepackage{graphicx}    

 \usepackage{rotating}    

 \allowdisplaybreaks[4]   

\begin{document}

 \newcommand{\be}[1]{\begin{equation}\label{#1}}
 \newcommand{\ee}{\end{equation}}
 \newcommand{\bea}{\begin{eqnarray}}
 \newcommand{\eea}{\end{eqnarray}}
 \def\disp{\displaystyle}

 \newcommand{\dmunit}{ {\rm pc\hspace{0.24em} cm^{-3}} }
 \newcommand{\cmn}{,\,}

 \begin{titlepage}

 \begin{flushright}
 arXiv:2408.12983
 \end{flushright}

 \title{\Large \bf Checking the Empirical Relations with the Current\\
 Localized Fast Radio Bursts}

 \author{Lin-Yu~Li\,$^{a,}$\footnote{email
 address:\ lilinyu0814@163.com}\,,
 Jing-Yi~Jia\,$^a$\,,
 Da-Chun~Qiang\,$^b$\,,
 Hao~Wei\,$^{a,}$\footnote{Corresponding author;\ email
 address:\ haowei@bit.edu.cn}\vspace{2.4mm}}
 \affiliation{$^{a)\,}$School of Physics, Beijing
 Institute of Technology, Beijing 100081, China\vspace{2mm}\\
 $^{b)\,}$Institute for Gravitational Wave Astronomy, Henan Academy of
 Sciences, Zhengzhou 450046, Henan, China}

 \begin{abstract}\vspace{1cm}
 \centerline{\bf ABSTRACT}\vspace{2mm}
 Although fast radio bursts (FRBs) were discovered more than a decade
 ago, and they have been one of the active fields in astronomy
 and cosmology, their origins are still unknown.~An interesting
 topic closely related to the origins of FRBs is their
 classifications.~Different classes of FRBs require different physical
 mechanisms.~If some empirical relations are found for different classes
 of FRBs, they might justify the classifications scenario and help us to
 reveal the physical mechanisms behind. On the other hand, FRBs are
 actually a promising probe for cosmology, since their redshifts could
 be $z\sim 3$ or even higher.~Similar to the cosmology of type Ia supernovae
 (SNIa) or Gamma-ray bursts (GRBs), some empirical relations might also play
 an important role in the FRB cosmology.~In the literature, some new
 classifications of FRBs different from repeaters and non-repeaters
 were proposed recently.~In particular, it was suggested to classify
 FRBs into the ones associated with old or young stellar populations,
 and some empirical relations have also been found for them,
 respectively.~One of these empirical relations (namely $L_\nu-E$ relation)
 without dispersion measure (DM) has been used to calibrate FRBs as
 standard candles for cosmology.~This shows the potential of the new
 classification and the empirical relations for FRBs.~Nowadays,
 more than 50 FRBs have been well localized, and hence their redshifts
 $z$ are observationally known.~So, it is of interest to check the empirical
 relations with the actual data of current localized FRBs.~We find that
 many empirical relations still hold, and in particular the one used to
 calibrate FRBs as standard candles for cosmology stands firm.~This
 is beneficial to the FRB cosmology.
 \end{abstract}


 \maketitle

 \end{titlepage}

 \renewcommand{\baselinestretch}{1.0}


\section{Introduction}\label{sec1}

In the past decade, fast radio bursts (FRBs)~\cite{NAFRBs,
 Lorimer:2018rwi,Keane:2018jqo,Petroff:2021wug,Xiao:2021omr,
 Zhang:2020qgp,Zhang:2022uzl,Nicastro:2021cxs} have been an active field
 in astronomy and cosmology. One of the key measured quantities of FRBs
 is the dispersion measure (DM), which is usually large and well in
 excess of the Galactic values. Since almost all of FRBs are at
 extragalactic/cosmological distances, they are actually a promising
 probe to study cosmology and the intergalactic medium (IGM). As a very
 crude rule of thumb, the redshift of FRB
 $z\sim {\rm DM}/(1000\;\dmunit)$~\cite{Lorimer:2018rwi}.~Note that
 the current observed FRBs reach up to $\rm DM\gtrsim 3000\;\dmunit$
 (see e.g.~\cite{tns,Spanakis-Misirlis:2022,
 frbstats,Xu:2023did,blinkverse}), and hence the inferred redshifts
 could be $z\sim 3$. Actually, it is expected that FRBs are detectable
 up to redshift $z\sim 15$ in e.g.~\cite{Zhang:2018csb}.~So, FRBs could
 be a useful probe for cosmology, complementarily to e.g. type
 Ia supernovae (SNIa) and cosmic microwave background (CMB).

Although FRBs have been discovered more than a decade ago, their origins
 are still unknown~\cite{NAFRBs,Lorimer:2018rwi,Keane:2018jqo,
 Petroff:2021wug,Xiao:2021omr,Zhang:2020qgp,Zhang:2022uzl,Nicastro:2021cxs}.
 To this end, a lot of theoretical models were proposed in the
 literature (see e.g.~the living FRB theory
 catalog~\cite{Platts:2018hiy,frbtheorycat}).~On the other hand, the
 observational data of FRBs were rapidly accumulated in the
 recent years, which are fairly helpful to study e.g.~the engines, radiation
 mechanisms, propagation effects, distributions, classifications, and
 cosmological applications of FRBs.

An interesting topic closely related to the origins of FRBs is their
 classifications~\cite{Caleb:2018ygr}. It is very natural to ask that
 ``\;how many different populations of FRBs exist?\;''~Actually, it was
 strongly argued in e.g.~\cite{Palaniswamy:2017aze,Zhong:2022uvu} that a
 single population cannot account for the observational data of
 FRBs.~Obviously, different physical mechanisms for the origins
 of FRBs are required by different classes of FRBs. In the
 literature~\cite{Petroff:2021wug,Xiao:2021omr,Zhang:2020qgp,Zhang:2022uzl,
 Nicastro:2021cxs,Caleb:2018ygr}, the well-known classification
 is repeating and non-repeating FRBs.~Note that the repeaters rule out
 the cataclysmic origins for these sources. However, the question is
 whether the (apparent) non-repeaters are genuinely one-off or not,
 and it is still  extensively debated in the literature.

On the other hand, some new classifications of FRBs different from repeaters
 and non-repeaters have also been suggested in the literature.~For example,
 similar to the well-known classification of Gamma-ray bursts (GRBs), it
 was proposed in~\cite{Li:2021yds} to classify FRBs into short
 ($W<100\,\rm ms$) and long ($W>100\,\rm ms$) bursts, where $W$ is the pulse
 width.~A~tight power-law correlation between fluence and peak flux density
 was found for them.~In~\cite{Xiao:2021viy}, the repeating FRBs were
 classified into classical ($T_B\geq 10^{33}\,\rm K$) and
 atypical ($T_B<10^{33}\,\rm K$) bursts, where $T_B$ is the brightness
 temperature.~A~tight power-law correlation between pulse width and fluence
 was also found for classical bursts.~In~\cite{Chaikova:2022vnh}, two major
 classes of FRBs featuring different waveform morphologies and
 simultaneously different distributions of brightness temperature $T_B$
 were identified. In the literature, the relevant works based on machine
 learning are notable.

In~\cite{Guo:2022wpf}, it was suggested to classify FRBs into the ones
 associated with old or young stellar populations. One can call
 them Classes (a) and (b) FRBs as in~\cite{Guo:2022wpf,Guo:2023hgb},
 or alternatively, oFRBs and yFRBs as in the present work (see below,
 n.b.~Table~\ref{tab1}). In fact, this new classification of FRBs is
 similar to the well-known classification of type I GRBs (typically
 short and associated with old populations) and type II GRBs
 (typically long and associated with young populations)
 \cite{Zhang:2006mb,Kumar:2014upa}. The Galactic FRB 200428 associated
 with the young magnetar SGR 1935+2154 \cite{Andersen:2020hvz,
 Bochenek:2020zxn,Lin:2020mpw,Li:2020qak} confirmed that some
 FRBs originate from young magnetars. So, it is reasonable to expect
 that the yFRB distribution is closely correlated with star-forming
 activities. To date, a lot of FRBs are well localized to the
 star-forming galaxies (or even the star-forming regions in their host
 galaxies).~We present these yFRBs associated with young stellar populations
 in Table~\ref{tab2}. On the other hand, the well-known
 repeating FRB 20200120E in a globular cluster of the nearby galaxy
 M81~\cite{Bhardwaj:2021xaa,Kirsten:2021llv,Nimmo:2021yob} suggested
 that some FRBs are associated with old stellar populations
 instead.~Nowadays, some FRBs are well localized to old galaxies with
 low star formation rate (or even old regions in their host
 galaxies).~We also present these oFRBs associated with old
 stellar populations in Table~\ref{tab2}.

In addition, it was claimed in~\cite{Zhang:2021kdu} that the
 bursts of the first CHIME/FRB catalog~\cite{CHIMEFRB:2021srp,
 CHIMEFRB:cat1} as a whole do not track the cosmic star
 formation history (SFH). In~\cite{Qiang:2021ljr}, it was independently
 confirmed that the FRB distribution tracking SFH can be rejected at
 high confidence, and a suppressed evolution (delay) with respect to
 SFH was found. Thus, it cannot be true that all FRBs are associated
 with young stellar populations. There must be FRB progenitor
 formation channels associated with old stellar populations. So, the
 new classification of FRBs mentioned above, namely Classes (a) and (b)
 FRBs as in~\cite{Guo:2022wpf,Guo:2023hgb} (or alternatively, oFRBs and
 yFRBs as in the present work), has been well motivated.

If the existing classification non-repeaters/repeaters is valid (namely
 there are genuinely one-off FRBs), one might combine these two
 classifications to form a new subclassification of FRBs. We present a
 brief summary of the universal subclassification scheme of FRBs in
 Table~\ref{tab1}. One can call them Type I, II, a, b, Ia, Ib, IIa, IIb
 FRBs as in~\cite{Guo:2022wpf,Guo:2023hgb}, respectively.~However, these
 terms might be hard to remember.~Note that the term ``\,rFRBs\,'' has
 been strongly suggested for repeating FRBs in e.g.~\cite{Zhang:2022uzl,
 Nicastro:2021cxs,Yamasaki:2017hdr,Ouyed:2019ikb,Pilia:2020ork}.~On the
 other hand, in the field of molecular biology, some subclasses of the
 well-known RNA (ribonucleic acid)~\cite{RNA} and DNA (deoxyribonucleic
 acid)~\cite{DNA} are called e.g.~mRNA, tRNA, rRNA and tmRNA, miRNA,
 snRNA, crRNA, sgRNA, as well as eDNA, gDNA, cDNA and dsDNA, ssDNA,
 cfDNA. Therefore, one can also use the new terms nFRBs, rFRBs, oFRBs,
 yFRBs, noFRBs, nyFRBs, roFRBs, ryFRBs alternatively, as shown
 in Table~\ref{tab1}, while the suffix ``\,s\,'' is just for the plural
 and it can be removed for the singular. They are friendly and easy to
 remember in fact.

Notice that even if the existing classification non-repeaters/repeaters
 is invalid (namely there are no genuinely one-off FRBs, and all FRBs
 repeat), the new classification oFRBs/yFRBs still holds well
 by itself.~Please keep this in mind carefully.


\begin{table}[tb]
 \renewcommand{\arraystretch}{1.0}
 \begin{center}
 \vspace{-3mm}  
 \hspace{-4.5mm}  
 \begin{tabular}{c|c|c} \hline\hline
 {\bf FRBs} & \begin{tabular}{c} $\vspace{-3.7mm}$\\ {\bf Class (a)\;:} \\ associated with old stellar populations \\ {\bf (\,oFRBs\,)} \\[1.295mm] \end{tabular}
 & \begin{tabular}{c} {\bf Class (b)\;:} \\ \ associated with young stellar populations \\ {\bf (\,yFRBs\,)} \end{tabular} \\ \hline
 \begin{tabular}{c} {\bf Type I\;:} \\ Non-repeating\ \ \\ {\bf (\,nFRBs\,)} \\ \end{tabular}
 & \begin{tabular}{c} $\vspace{-3.7mm}$\\ {\bf Type Ia\;:} \\ Non-repeating FRBs \\ \ associated with old stellar populations \ \\ {\bf (\,noFRBs\,)} \\[1.295mm] \end{tabular}
 & \begin{tabular}{c} {\bf Type Ib\;:} \\ Non-repeating FRBs \\ \ associated with young stellar populations\ \\ {\bf (\,nyFRBs\,)} \end{tabular} \\ \hline
 \begin{tabular}{c} {\bf Type II\;:} \\ Repeating\ \ \\ {\bf (\,rFRBs\,)} \end{tabular}
 & \begin{tabular}{c} $\vspace{-3.7mm}$\\ {\bf Type IIa\;:} \\ Repeating FRBs \\ \ associated with old stellar populations \ \\ {\bf (\,roFRBs\,)} \\[1.295mm] \end{tabular}
 & \begin{tabular}{c} {\bf Type IIb\;:} \\ Repeating FRBs \\ \ associated with young stellar populations \\ {\bf (\,ryFRBs\,)} \\ \end{tabular} \\ \hline \hline
 \end{tabular}
 \end{center}
 \vspace{-2mm}  
 \caption{\label{tab1} A brief summary of the universal
 subclassification scheme of FRBs proposed in~\cite{Guo:2022wpf}. One
 can call them Type I, II, a, b, Ia, Ib, IIa, IIb FRBs as
 in~\cite{Guo:2022wpf,Guo:2023hgb}, or alternatively, nFRBs, rFRBs,
 oFRBs, yFRBs, noFRBs, nyFRBs, roFRBs, ryFRBs as in the present work,
 respectively. See Sec.~\ref{sec1} for details.}
 \end{table}


If some empirical relations are found for different classes of FRBs,
 they might justify the classifications scenario and help us to
 reveal the physical mechanisms behind.~Actually, we have found
 some empirical relations in~\cite{Guo:2022wpf} for nFRBs in the first
 CHIME/FRB catalog~\cite{CHIMEFRB:2021srp,CHIMEFRB:cat1}, e.g.~the empirical
 relations between spectral luminosity $L_\nu$\,, isotropic energy $E$
 and $\rm DM_E$ (extragalactic DM), where
 \bea
 & L_\nu=4\pi d_L^2 S_\nu\,,\quad\quad E=4\pi
 d_L^2 \nu_c\, F_\nu/(1+z)\,,\label{eq1}\\[1mm]
 & {\rm DM_E\equiv DM_{obs}-DM_{MW}-DM_{halo}
 =DM_{IGM}+DM_{host}}/(1+z)\,,\label{eq2}
 \eea
 in which $d_L$ is the luminosity distance, $S_\nu$ is the
 flux, $F_\nu$ is the specific fluence, $\nu_c$ is the central
 observing frequency, $\rm DM_{obs}$ is the observed DM, and
 $\rm DM_{MW}$, $\rm DM_{halo}$, $\rm DM_{IGM}$, $\rm DM_{host}$ are
 the contributions from the Milky Way (MW), the MW halo, IGM, the host
 galaxy (including interstellar medium of the host galaxy and
 the near-source plasma), respectively. In~\cite{Guo:2022wpf},
 the tight 2D empirical relations for nyFRBs in the first
 CHIME/FRB catalog are found to be
 \bea
 &&\log E=0.8862\,\log L_\nu+10.0664\,,\label{eq3}\\[1mm]
 &&\log L_\nu=2.4707\,\log {\rm DM_E}+27.3976\,,\label{eq4}\\[1mm]
 &&\log E=2.2345\,\log {\rm DM_E}+34.2238\,,\label{eq5}
 \eea
 where ``\,$\log$\,'' gives the logarithm to base $10$, and
 $E$, $L_\nu$\,, $\rm DM_E$ are in units of $\rm erg$, $\rm erg/s/Hz$,
 $\dmunit$, respectively. There are similar 2D empirical relations for
 noFRBs and all nFRBs (\,= noFRBs + nyFRBs) in the first
 CHIME/FRB catalog, but with quite different slopes and intercepts.
 On the other hand, some tight 2D empirical relations are found only
 for noFRBs in the first CHIME/FRB catalog, namely~\cite{Guo:2022wpf}
 \bea
 &&\log F_\nu=0.7709\,\log S_\nu+0.3150\,,\label{eq6}\\[1mm]
 &&\log {\rm DM_E}=-0.3987\,\log S_\nu+2.5402\,,\label{eq7}\\[1mm]
 &&\log F_\nu=-1.2873\,\log {\rm DM_E}+3.6139\,,\label{eq8}
 \eea
 where $F_\nu$, $S_\nu$, $\rm DM_E$ are in units of $\rm Jy\,ms$, $\rm
 Jy$, $\dmunit$, respectively. It is worth noting that there are no such
 empirical relations for nyFRBs in the first CHIME/FRB catalog.~Finally,
 some tight 3D empirical relations are also found.~We refer to Sec.~4
 of~\cite{Guo:2022wpf} for details.

Similar to the cosmology of SNIa or GRBs, the empirical relations might also
 play an important role in the FRB cosmology. In~\cite{Guo:2023hgb},
 the empirical $L_\nu - E$ relation for yFRBs akin to Eq.~(\ref{eq3}),
 namely
 \be{eq9}
 \log\frac{E}{\rm erg}=a\log\frac{L_\nu}{\rm erg/s/Hz}+b\,,
 \ee
 has been used to calibrate yFRBs as standard candles for cosmology,
 where slope $a$ and intercept $b$ are both dimensionless constants.~In this
 way, one can study cosmology by using the luminosity distances $d_L$ of
 yFRBs directly (rather than DM as an indirect proxy of $d_L$), which could
 be obtained from Eq.~(\ref{eq9}) if $a\not=1$, noting that $d_L$ exists
 implicitly in both $L_\nu$ and $E$ (n.b.~Eq.~(\ref{eq1})).~Actually, as
 shown in~\cite{Guo:2023hgb} by simulations, this method works well.~Noting
 that DM is not involved in the empirical $L_\nu - E$ relation, one can
 avoid the large uncertainties of $\rm DM_{IGM}$ and $\rm DM_{host}$ in
 DM which plague the FRB cosmology. This shows the potential of the
 new classification yFRBs/oFRBs and the empirical relations for them.

However, it is well known that most of FRBs in the first
 CHIME/FRB catalog~\cite{CHIMEFRB:2021srp,CHIMEFRB:cat1}
 are not localized, and hence their redshifts $z$ and the corresponding
 luminosity distances $d_L$ are unknown actually. In finding
 the empirical relations mentioned above, the redshifts $z$ of FRBs
 in~\cite{Guo:2022wpf} were inferred from DMs, and hence they are not
 real redshifts measured directly. Thus, it is natural to ask
 ``\,whether these empirical relations for FRBs are real or not?\,''~Several
 years passed after the first CHIME/FRB catalog \cite{CHIMEFRB:2021srp,
 CHIMEFRB:cat1}.~Nowadays, more than 50 FRBs have been well localized,
 and hence their redshifts $z$ are observationally known.~So, it is of
 interest to check~the empirical relations with the current localized FRBs.

The rest of this paper is organized as follows.~In Sec.~\ref{sec2}, we
 briefly introduce the sample of current localized FRBs.~In Sec.~\ref{sec3},
 we consider the empirical relations for the current localized FRBs.~In
 Sec.~\ref{sec4}, the uncertainties are also taken into account.~In
 Sec.~\ref{secins}, we further discuss the independencies of empirical
 relations.~In Sec.~\ref{sec5}, some brief concluding remarks are given.


 \begin{sidewaystable}[p] 
 \renewcommand{\arraystretch}{1.7}
 \begin{center}
 \vspace{-3mm}  
 \hspace{-4mm}  
 \begin{tabular}{ccccccccccccccc} \hline\hline
 FRB & R.A. & Dec. & Redshift & $\rm DM_{obs}$ & Width & Fluence & Flux & $\nu_p$ & $\nu_c$ & $\Delta\nu$ & Rep. & Pop. & Telescope & \ Ref.\ \\[-2mm]
 (name) & (deg.) & \ (deg.)\ & $z$ & (\,$\dmunit$\,) & (ms) & (Jy\,ms) & (Jy) & (MHz) & (MHz) & (MHz) & (0/1) & (0/1/2) & \\ \hline
 20220207C & 310.1995 & 72.8823 & \;0.04304\; & $262.38\pm 0.01$ & 0.5 & 16.2 & & & 1405 & 187.5 & 0 & 1 & DSA-110 & \cite{Law:2023ibd} \\[-1.5mm]
 20220307B & 350.8745 & 72.1924 & 0.248123 & $499.27\pm 0.06$ & 0.5 & 3.2 & & & 1405 & 187.5 & 0 & 1 & DSA-110 & \cite{Law:2023ibd} \\[-1.5mm]
 20220310F & 134.7204 & 73.4908 & \ 0.477958 \ & $462.24\pm 0.005$ & 1.0 & 26.2 & & & 1405 & 187.5 & 0 & 1 & DSA-110 & \cite{Law:2023ibd} \\[-1.5mm]
 20220319D & 32.1779 & 71.0353 & 0.011228 & $110.98\pm 0.02$ & 0.3 & 8.0 & & & 1405 & 187.5 & 0 & 1 & DSA-110 & \cite{Law:2023ibd} \\[-1.5mm]
 20220418A & 219.1056 & 70.0959 & 0.622 & $623.25\pm 0.01$ & 1.0 & 4.2 & & & 1405 & 187.5 & 0 & 1 & DSA-110 & \cite{Law:2023ibd} \\[-1.5mm]
 20220506D & 318.0448 & 72.8273 & 0.30039 & $396.97\pm 0.02$ & 0.5 & 13.2 & & & 1405 & 187.5 & 0 & 1 & DSA-110 & \cite{Law:2023ibd} \\[-1.5mm]
 20220509G & 282.67 & 70.2438 & 0.0894 & $269.53\pm 0.02$ & 0.5 & 5.8 & & & 1405 & 187.5 & 0 & 0 & DSA-110 & \cite{Law:2023ibd} \\[-1.5mm]
 20220825A & 311.9815 & 72.5850 & 0.241397 & $651.24\pm 0.06$ & 1.0 & 5.8 & & & 1405 & 187.5 & 0 & 1 & DSA-110 & \cite{Law:2023ibd} \\[-1.5mm]
 20220914A & 282.0568 & 73.3369 & 0.1139 & $631.28\pm 0.04$ & 0.5 & 2.6 & & & 1405 & 187.5 & 0 & 1 & DSA-110 & \cite{Law:2023ibd} \\[-1.5mm]
 20220920A & 240.2571 & 70.9188 & 0.158239 & $314.99\pm 0.01$ & 0.5 & 3.9 & & & 1405 & 187.5 & 0 & 1 & DSA-110 & \cite{Law:2023ibd} \\[-1.5mm]
 20221012A & 280.7987 & 70.5242 & 0.284669 & $441.08\pm 0.7$ & 2.0 & 5.1 & & & 1405 & 187.5 & 0 & 0 & DSA-110 & \cite{Law:2023ibd} \\[-1.5mm]
 20220912A & 347.2704 & 48.7071 & 0.0771 & $219.46\pm 0.042$ & 4.7 & $0.552\pm 0.007$ & \ $0.1174\pm 0.0014$ \ & 1479.5 & 1250 & 500 & 1 & 1 & FAST & \cite{DeepSynopticArrayTeam:2022rbq,Zhang:2023eui} \\[-1.5mm]
 20210117A & 339.9792 & $-16.1515$ & 0.214 & $729.1^{+0.36}_{-0.23}$ & $0.14\pm 0.01$ & $36^{+28}_{-9}$ & & & 1271.5 & 336 & 0 & 2 & ASKAP & \cite{Bhandari:2022ton} \\[-1.5mm]
 20181220A & 348.6982 & 48.3421 & 0.02746 & $208.66\pm 1.62$ & 2.95 & $3.0\pm 1.7$ & $1.33\pm 0.82$ & 400.2 & 600 & 400 & 0 & 1 & CHIME & \cite{Bhardwaj:2023vha,CHIMEFRB:2021srp,CHIMEFRB:cat1} \\[-1.5mm]
 20181223C & 180.9207 & 27.5476 & 0.03024 & $111.61\pm 1.62$ & 1.97 & $2.84\pm 0.93$ & $1.36\pm 0.51$ & 479.2 & 600 & 400 & 0 & 1 & CHIME & \cite{Bhardwaj:2023vha,CHIMEFRB:2021srp,CHIMEFRB:cat1} \\[-1.5mm]
 20190418A & 65.8123 & 16.0738 & 0.07132 & $182.78\pm 1.62$ & 1.97 & $2.2\pm 1.0$ & $0.99\pm 0.58$ & 400.2 & 600 & 400 & 0 & 1 & CHIME & \cite{Bhardwaj:2023vha,CHIMEFRB:2021srp,CHIMEFRB:cat1} \\[-1.5mm]
 20190425A & 255.6625 & 21.5767 & 0.03122 & $127.78\pm 1.62$ & 0.98 & $31.6\pm 4.2$ & $18.6\pm 2.6$ & 591.8 & 600 & 400 & 0 & 1 & CHIME & \cite{Bhardwaj:2023vha,CHIMEFRB:2021srp,CHIMEFRB:cat1} \\[-1.5mm]
 20220610A & 351.0732 & $-33.5137$ & 1.016 & $1458.15^{+0.25}_{-0.55}$ & $0.41\pm 0.01$ & $45\pm 5$ & & & 1271.5 & 336 & 0 & 1 & ASKAP & \cite{Ryder:2022qpg} \\[-1.5mm]
 20200120E & \ 149.4863 \ & 68.8256 & $-0.0001$ & $87.782\pm 0.003$ & \;$0.16\pm 0.05$\; & $2.25\pm 0.12$ & $1.8\pm 0.9$ & & 600 & 400 & 1 & 0 & CHIME & \cite{Bhardwaj:2021xaa} \\[-1.5mm]
 20171020A & 333.75 & $-19.6667$ & 0.008672 & $114.1\pm 0.2$ & $3.2\pm 3.2$ & $200\pm 500$ & 117.6 & & 1297 & 336 & 0 & 1 & ASKAP & \cite{Mahony:2018ddp} \\[-1.5mm]
 20121102A & 82.9946 & 33.1479 & 0.1927 & $557\pm 2$ & $3.0\pm 0.5$ & 1.2 & $0.4\pm 0.1$ & 1600 & 1375 & 322.6 & 1 & 1 & Arecibo & \cite{Gordon:2023cgw} \\[-1.5mm]
 20180301A & 93.2268 & 4.6711 & 0.3304 & $536^{+8}_{-13}$ & $2.18\pm 0.06$ & 1.3 & $1.2\pm 0.1$ & 1415 & 1352 & 338.281 & 1 & 1 & Parkes & \cite{Gordon:2023cgw,Price:2019fmc} \\[-1.5mm]
 20180916B & 29.5031 & 65.7168 & 0.0337 & $347.8\pm 0.0058$ & 3.93 & $6.1\pm 1.7$ & $1.84\pm 0.72$ & 603.9 & 600 & 400 & 1 & 2 & CHIME & \cite{Gordon:2023cgw,CHIMEFRB:2021srp,CHIMEFRB:cat1} \\[-1.5mm]
 20180924B & 326.1053 & $-40.90$ & 0.3212 & $362.42\pm 0.06$ & $1.3\pm 0.09$ & $16\pm 1$ & 12.3 & & 1320 & 336 & 0 & 1 & ASKAP & \cite{Gordon:2023cgw} \\[-1.5mm]
 20181112A & 327.3485 & $-52.9709$ & 0.4755 & $589.27\pm 0.03$ & $2.1\pm 0.2$ & $26\pm 3$ & & & 1272.5 & 336 & 0 & 1 & ASKAP & \cite{Gordon:2023cgw} \\
 \hline\hline
 \end{tabular}
 \end{center}
 \vspace{-2mm}  
 \caption{\label{tab2} The sample of current 52 localized
 FRBs.~Right Ascension~(R.A.)~and Declination~(Dec.)~are given
 in degree,~J2000.~$\nu_p$, $\nu_c$ and $\Delta\nu$ are peak frequency,
 central frequency and bandwidth, respectively. In the 12th column (Rep.), 0
 and 1 indicate non-repeaters and repeaters, respectively. In the 13th
 column (Pop.), 0, 1 and 2 indicate FRBs associated with old, young
 and unknown/transitional stellar populations, respectively. See the
 text for details.}
 \end{sidewaystable}



 \begin{sidewaystable}[p] 
 \renewcommand{\arraystretch}{1.7}
 \begin{center}
 \vspace{-2.4mm}  
 \hspace{-4mm}  
 \begin{tabular}{ccccccccccccccc} \hline\hline
 FRB & R.A. & Dec. & Redshift & $\rm DM_{obs}$ & Width & Fluence & Flux & $\nu_p$ & $\nu_c$ & $\Delta\nu$ & Rep. & Pop. & Telescope & \ Ref.\ \\[-2mm]
 (name) & (deg.) & \ (deg.)\ & $z$ & (\,$\dmunit$\,) & (ms) & (Jy\,ms) & (Jy) & (MHz) & (MHz) & (MHz) & (0/1) & (0/1/2) & \\ \hline
 20190102C & \ 322.4157 \ & $-79.4757$ & 0.2912 & $363.6\pm 0.3$ & $1.7\pm 0.1$ & $14\pm 1$ & & & 1320 & 336 & 0 & 1 & ASKAP & \cite{Gordon:2023cgw} \\[-1.5mm]
 20190520B & 240.5178 & $-11.2881$ & 0.2414 & $1201\pm 10$ & $8.7\pm 0.8$ & $0.075\pm 0.003$ & & & 1375 & 400 & 1 & 1 & FAST & \cite{Gordon:2023cgw} \\[-1.5mm]
 20190608B & 334.0199 & $-7.8983$ & 0.1178 & $338.7\pm 0.5$ & $6.0\pm 0.8$ & $26\pm 4$ & 4.3 & 1295 & 1320 & 336 & 0 & 1 & ASKAP & \cite{Gordon:2023cgw,Hiramatsu:2022tyn} \\[-1.5mm]
 20190611B & 320.7456 & $-79.3976$ & 0.3778 & $321.4\pm 0.2$ & $2\pm 1$ & $10\pm 2$ & & 1152 & 1320 & 336 & 0 & 1 & ASKAP & \cite{Gordon:2023cgw} \\[-1.5mm]
 20190711A & 329.4193 & $-80.358$ & 0.522 & $593.1\pm 0.4$ & $6.5\pm 0.5$ & $34\pm 3$ & & 1152 & 1320 & 336 & 1 & 1 & ASKAP & \cite{Gordon:2023cgw,Macquart:2020lln} \\[-1.5mm]
 20190714A & 183.9797 & $-13.021$ & 0.2365 & $504.13\pm 2.0$ & 1 & 8 & 8 & 1272.5 & 1272.5 & 336 & 0 & 1 & ASKAP & \cite{Gordon:2023cgw,Hiramatsu:2022tyn,HESS:2021smp,Guidorzi:2020ggq} \\[-1.5mm]
 20191001A & 323.3513 & $-54.7478$ & 0.234 & $506.92\pm 0.04$ & $0.22\pm 0.03$ & $143\pm 15$ & & 1088 & 920.5 & 336 & 0 & 1 & ASKAP & \cite{Gordon:2023cgw,Bhandari:2020cde} \\[-1.5mm]
 20200430A & 229.7064 & 12.3768 & 0.1608 & $380.1\pm 0.4$ & & $35\pm 4$ & & 864.5 & 864.5 & 336 & 0 & 1 & ASKAP & \cite{Gordon:2023cgw,Hiramatsu:2022tyn} \\[-1.5mm]
 20200906A & 53.4962 & $-14.0832$ & 0.3688 & $577.8\pm 0.2$ & $6.0\pm 0.2$ & $59^{+25}_{-10}$ & 9.8 & 864.5 & 864.5 & 336 & 0 & 1 & ASKAP & \cite{Gordon:2023cgw,Hiramatsu:2022tyn} \\[-1.5mm]
 20201124A & 77.0146 & 26.0607 & 0.0979 & $415.3\pm 0.63$ & $22\pm 1$ & $6\pm 2$ & $0.8\pm 0.5$ & 668 & 600 & 400 & 1 & 1 & CHIME & \cite{Gordon:2023cgw,Lanman:2021yba} \\[-1.5mm]
 20210320C & 204.4608 & $-16.1227$ & 0.2797 & $384.8\pm 0.3$ & & & & & 864.5 & 336 & 0 & 1 & ASKAP & \cite{Gordon:2023cgw} \\[-1.5mm]
 20210410D & 326.0863 & $-79.3182$ & 0.1415 & $571.2\pm 1.0$ & & 35.4 & 1.5 & 1711.58 & 1284 & 856 & 0 & 2 & MeerKAT & \cite{Gordon:2023cgw,Caleb:2023atr} \\[-1.5mm]
 20210807D & 299.2214 & $-0.7624$ & 0.1293 & $251.9\pm 0.2$ & & $113\pm 9$ & & & 920.5 & 336 & 0 & 0 & ASKAP & \cite{Gordon:2023cgw} \\[-1.5mm]
 20211127I & 199.8082 & $-18.8378$ & 0.0469 & $234.83\pm 0.08$ & 1.182 & $31\pm 1$ & & 1271.5 & 1271.5 & 336 & 0 & 1 & ASKAP & \cite{Gordon:2023cgw} \\[-1.5mm]
 20211203C & 204.5625 & $-31.3801$ & 0.3439 & $636.2\pm 0.4$ & & $28\pm 2$ & & & 920.5 & 336 & 0 & 1 & ASKAP & \cite{Gordon:2023cgw} \\[-1.5mm]
 20211212A & 157.3509 & 1.3609 & 0.0707 & $206\pm 5$ & & & & & 1631.5 & 336 & 0 & 1 & ASKAP & \cite{Gordon:2023cgw} \\[-1.5mm]
 20220105A & 208.8039 & 22.4665 & 0.2785 & $583\pm 1$ & & & & & 1631.5 & 336 & 0 & 1 & ASKAP & \cite{Gordon:2023cgw} \\[-1.5mm]
 20191106C & 199.5801 & 42.9997 & 0.10775 & $332.2\pm 0.7$ & 10.81 & $1.48\pm 0.26$ & & 618.5 & 600 & 400 & 1 & 1 & CHIME & \cite{Ibik:2023ugl,CHIMEFRB:2021srp,CHIMEFRB:cat1} \\[-1.5mm]
 20200223B & 8.2695 & 28.8313 & 0.06024 & $201.8\pm 0.4$ & 3.93 & $1.06\pm 0.36$ & & 754.9 & 600 & 400 & 1 & 1 & CHIME & \cite{Ibik:2023ugl,CHIMEFRB:2021srp,CHIMEFRB:cat1} \\[-1.5mm]
 20190110C & 249.3185 & 41.4434 & 0.12244 & $221.6\pm 1.6$ & 0.39 & $1.4\pm 0.76$ & $0.64\pm 0.39$ & 427.4 & 600 & 400 & 1 & 1 & CHIME & \cite{Ibik:2023ugl,CHIMEFRB:2021srp,CHIMEFRB:cat1} \\[-1.5mm]
 20190303A & 207.9958 & 48.1211 & 0.064 & $223.2\pm 0.017$ & 3.93 & $2.54\pm 0.97$ & $0.47\pm 0.26$ & 631.5 & 600 & 400 & 1 & 1 & CHIME & \cite{Michilli:2022bbs,CHIMEFRB:2021srp,CHIMEFRB:cat1} \\[-1.5mm]
 20180814A & 65.6833 & 73.6644 & 0.068 & $190.9\pm 0.076$ & 7.86 & $2.6\pm 1.0$ & $0.39\pm 0.3$ & 464.2 & 600 & 400 & 1 & 0 & CHIME & \cite{Michilli:2022bbs,CHIMEFRB:2021srp,CHIMEFRB:cat1} \\[-1.5mm]
 20210405I & 255.3397 & $-49.5451$ & 0.066 & $565.17\pm 0.49$ & \;$8.67\pm 0.28$\; & 120.8 & 15.9 & 1016.5 & 1284 & 1712 & 0 & 0 & MeerKAT & \cite{Driessen:2023lxj} \\[-1.5mm]
 20191228A & 344.4304 & $-28.5941$ & 0.2432 & $297.5\pm 0.05$ & $2.3\pm 0.6$ & $40^{+100}_{-40}$ & 17 & 1271.5 & 1272.5 & 336 & 0 & 1 & ASKAP & \cite{Bhandari:2021pvj} \\[-1.5mm]
 20181030A & 158.5838 & 73.7514 & 0.00385 & $103.5\pm 0.3$ & 1.97 & $8.2\pm 5.9$ & $4.3\pm 3.6$ & 703.7 & 600 & 400 & 1 & 1 & CHIME & \cite{Bhardwaj:2021hgc,CHIMEFRB:2021srp,CHIMEFRB:cat1} \\[-1.5mm]
 20190523A & 207.065 & 72.4697 & 0.66 & $760.8\pm 0.6$ & $0.42\pm 0.05$ & 280 & 660 & 1530 & 1411 & 152.6 & 0 & 0 & DSA-110 & \cite{Ravi:2019alc} \\[-1.5mm]
 20190614D & 65.0755 & 73.7067 & 0.6 & $959.2\pm 5.0$ & 5 & $0.62\pm 0.07$ & $0.124\pm 0.014$ & & 600 & 400 & 0 & 2 & CHIME & \cite{Law:2020cnm,Hiramatsu:2022tyn} \\
 \hline\hline
 \end{tabular}
 \end{center}
 \vspace{-2mm}  
 \setcounter{table}{1}  
 \caption{\label{tab2b} Continued.}
 \end{sidewaystable}



\section{The sample of current localized FRBs}\label{sec2}

The sample of current localized FRBs used in this work is
 mainly compiled from DSA-110, ASKAP, CHIME/FRB and other telescopes.~We
 present them in Table~\ref{tab2}, which consists of 52 localized FRBs
 in total, while the references are also given.~Notice that the redshift
 of FRB 20200120E reads $-0.0001$, which is blueshift in fact, due to
 its peculiar velocity towards us. It is extremely close to us, so that
 it is decoupled from the cosmic expansion in fact.~On the other hand,
 the observed $\rm DM_{obs}$ of FRB 20220319D is much less than the
 $\rm DM_{MW}$ obtained from NE2001~\cite{Cordes:2002wz,Cordes:2003ik,
 Ocker:2024rmw,ne2001p,Price:2021gzo,pygedm} and YMW16~\cite{YMW16,
 Price:2021gzo,pygedm}, namely its $\rm DM_E<0$. Although it was shown
 in~\cite{Ravi:2023zfl} that uncertainties in NE2001 and YMW16 could
 still accommodate an extragalactic origin for FRB 20220319D,
 we consider that it is better to be conservative. So, we exclude FRBs
 20200120E and 20220319D from Table~\ref{tab2}, and 50 localized FRBs
 are left.

In principle, the fluence $F_\nu$ is an integral of flux $S_\nu$ with
 respect to time. If the pulse width $W$ of FRB is small enough
 ($W\sim {\cal O}(\rm ms)$ or smaller in fact), one has $F_\nu\simeq
 S_\nu W$. So, one could approximately estimate $\rm flux\simeq
 fluence/width$, as in the literature (e.g.~\cite{Petroff:2016tcr}).~In
 Table~\ref{tab2}, the observed values of the fluxes or the widths are
 absent for some FRBs, and they could be estimated from
 this approximation.~Unfortunately, for six FRBs in Table~\ref{tab2}
 (20200430A, 20210320C, 20210807D, 20211203C, 20211212A, 20220105A),
 at least two of width, flux, fluence are absent, and hence the
 estimation $\rm flux\simeq fluence/width$ cannot work. Thus, we drop
 them out, and then 44 localized FRBs are left in our sample.

Some derived quantities are required to check the empirical
 relations.~For a given FRB, $\rm DM_{MW}$ can be obtained by using
 NE2001~\cite{Cordes:2002wz,Cordes:2003ik,Ocker:2024rmw,
 ne2001p,Price:2021gzo,pygedm} with its Right Ascension (R.A.)
 and Declination (Dec.) from Table~\ref{tab2}. Following
 e.g.~\cite{Guo:2022wpf,Dolag:2014bca,Prochaska:2019mn}, we adopt $\rm
 DM_{halo}=30\,\dmunit$. So, its $\rm DM_E=DM_{obs}-DM_{MW}-DM_{halo}$
 is on hand.~Since the redshift $z$ of a localized FRB
 is observationally known, its luminosity distance $d_L$ is given by
 \vspace{-0.1mm} 
 \be{eq10}
 d_L=\left(1+z\right)d_C=\left(1+z\right)\frac{c}{H_0}
 \int_0^z\frac{d\tilde{z}}{h(\tilde{z})}\,,
 \quad\quad h(z)=\left[\Omega_m\left(1+z\right)^3
 +\left(1-\Omega_m\right)\right]^{1/2}\,,
 \ee
 where $d_C$ is the comoving distance, $c$ is the speed of light, $H_0$
 is the Hubble constant, and we adopt $\Omega_m=0.3153$,
 $H_0=67.36\;{\rm km/s/Mpc}$ from the Planck 2018
 results~\cite{Aghanim:2018eyx}. Thus, the spectral luminosity~$L_\nu$
 and isotropic energy $E$ can be obtained from Eq.~(\ref{eq1}). On the
 other hand, the brightness temperature $T_B$ is given by
 (e.g.~\cite{Guo:2022wpf,Xiao:2021viy,Pietka:2014wra,Majid:2021uli})
 \be{eq11}
 T_B=\frac{S_\nu\, d_L^2}{2\pi\kappa_B\left(\nu W\right)^2}
 =1.1\times 10^{35}\,{\rm K}\,\left(\frac{S_\nu}{\rm Jy}\right)
 \left(\frac{d_L}{\rm Gpc}\right)^2\left(\frac{\nu}{\rm
 GHz}\right)^{-2}\left(\frac{W}{\rm ms}\right)^{-2}\,,
 \ee
 where $\kappa_B$ is the Boltzmann constant.~Noting that the
 peak frequencies $\nu_p$ are absent for half of FRBs in
 Table~\ref{tab2}, we instead use $\nu$ as the central frequency $\nu_c$
 in Eq.~(\ref{eq11}).~In e.g.~\cite{Zhang:2022uzl,Luo:2022smj}, it was
 argued that the luminosity distance $d_L$ in $T_B$ should be replaced
 by the angular diameter distance $d_A=(1+z)^{-2}\,d_L$ with a factor
 $(1+z)^3$ due to cosmological effects, and hence the ``\,correct\,''
 brightness temperature $T_B^{\rm corr}$ has a $(1+z)$
 dependence (namely $T_B$ in Eq.~(\ref{eq11}) divided by
 $(1+z)$). However, as mentioned above, the main motivation
 of the present work is closely connected with the new classification
 oFRBs/yFRBs proposed in~\cite{Guo:2022wpf}, where the $\nu W-L_\nu$
 phase plane plays a very important role. In the (logarithmic)
 $\nu W-L_\nu$ phase plane (see Fig.~1 of~\cite{Guo:2022wpf}), the isotherms
 $T_B=const.$ are straight lines, which can be used to divide
 the $\nu W-L_\nu$ phase plane into regions dominated by oFRBs
 or yFRBs respectively (see Sec.~3 of~\cite{Guo:2022wpf} for
 details).~On the contrary, if one instead uses the ``\,correct\,''
 brightness temperature $T_B^{\rm corr}=T_B/(1+z)$ introduced
 in e.g.~\cite{Zhang:2022uzl,Luo:2022smj}, the isotherms
 $T_B^{\rm corr}=const.$ become irregular curves (actually it
 is difficult to properly draw them) in the $\nu W-L_\nu$ phase plane due to
 the $(1+z)$ dependence, and then the classification oFRBs/yFRBs cannot work
 well.~Thus, we intentionally use $T_B$ defined by Eq.~(\ref{eq11}) in
 this work, rather than $T_B^{\rm corr}=T_B/(1+z)$ introduced
 in e.g.~\cite{Zhang:2022uzl,Luo:2022smj}.~If necessary, one might call
 it the ``\,pseudo-brightness temperature\,'' (or something like that).

Since the conclusions of~\cite{Bhandari:2022ton,Gordon:2023cgw} on the
 star-forming activity of the host galaxy of FRB 20210117A are fairly
 different, we assign $\tt Pop.=2$ (namely unknown/transitional) to it
 in Table~\ref{tab2}. On the other hand, actually we have also compiled
 the spectral indices and other observational quantities for
 the localized FRBs. However, they are absent for most of FRBs
 in Table~\ref{tab2}, and no significant empirical correlations
 are found for them. So, we do not include them in Table~\ref{tab2}.


 \begin{center}
 \begin{figure}[tb]
 \centering
 \vspace{-6mm}  
 \hspace{-6mm}  
 \includegraphics[width=0.98\textwidth]{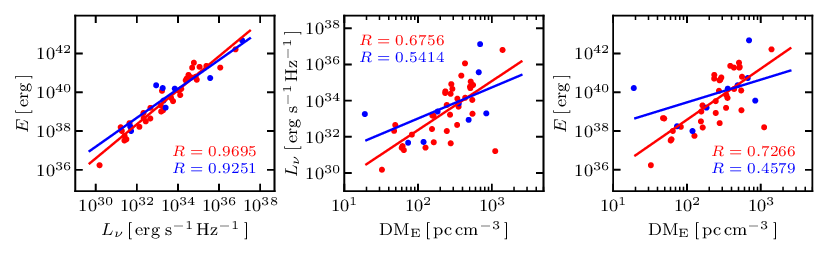}
 \vspace{-2mm}  
 \caption{\label{fig1} The empirical relations (red/blue
 lines) between spectral luminosity $L_\nu$, isotropic energy $E$ and
 $\rm DM_E$ for 35 localized yFRBs (red points) and 9 localized
 oFRBs (blue points), respectively.~The corresponding $R$ values (in
 red/blue) are also given. See Sec.~\ref{sec3a} for details.}
 \end{figure}
 \end{center}



 \begin{center}
 \begin{figure}[tb]
 \centering
 \vspace{-6mm}  
 \hspace{-6mm}  
 \includegraphics[width=0.98\textwidth]{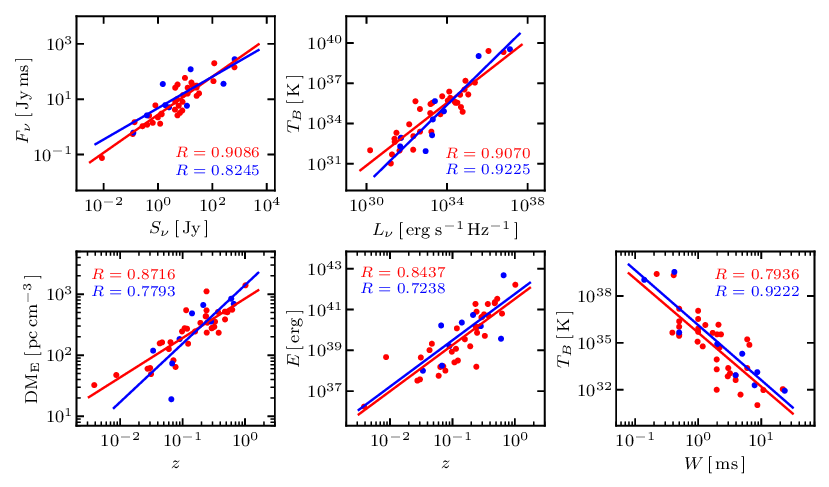}
 \vspace{-2mm}  
 \caption{\label{fig2} The other five empirical relations (red/blue
 lines) with high score $R^2 > 0.7$ for 35 localized yFRBs (red points)
 and 9 localized oFRBs (blue points), respectively.\hspace{1.28mm}The
 corresponding $R$ values (in red/blue) are also given. See
 Sec.~\ref{sec3a} for details.}
 \end{figure}
 \end{center}


\vspace{-17mm} 


\section{Empirical relations for the current localized FRBs}\label{sec3}

Since we have only 44 usable localized FRBs as mentioned in
 Sec.~\ref{sec2}, it is reasonable to consider the empirical
 relations just for the (large) classes yFRBs and oFRBs, or
 nFRBs and rFRBs, rather than the (small) subclasses nyFRBs, noFRBs,
 ryFRBs, roFRBs.

At first, we linearly fit the data points without error bars (and we
 will take uncertainties into account in Sec.~\ref{sec4}). This can
 be done by using {\tt sklearn.linear$_{-}$model.LinearRegression} in
 Python~\cite{LinearRegression}, which employs the ordinary
 least squares linear regression.~The score (coefficient of
 determination) is given by $R^2\equiv 1-\sum_k\left(y_k-
 \hat{y}_k\right)^2/\sum_k\left(y_k-\bar{y}\right)^2$, where $y_k$,
 $\hat{y}_k$ and $\bar{y}$ are the observed values, regressed
 values and mean of observed values~\cite{LinearRegression,Xiao:2021viy,
 Guo:2022wpf}, respectively.~The higher $R$ indicates the better fit,
 and $R=1$ at best.~For simplicity, we only consider 2D
 empirical relations in this work.


\subsection{Empirical relations for the localized
 yFRBs and oFRBs}\label{sec3a}

In~\cite{Guo:2022wpf}, some tight empirical relations for nyFRBs and
 noFRBs in the first CHIME/FRB catalog~\cite{CHIMEFRB:2021srp,CHIMEFRB:cat1}
 have been found, as mentioned in Sec.~\ref{sec1}. Here, we consider the
 empirical relations for 35 localized yFRBs (labeled with $\tt Pop.=1$)
 and 9 localized oFRBs (5 labeled with $\tt Pop.=0$, and 4 labeled with
 $\tt Pop.=2$, namely we also take FRBs associated with unknown/transitional
 stellar populations into account, since the number of oFRBs is too few)
 in Table~\ref{tab2}.

In Fig.~\ref{fig1}, we check the 2D empirical relations between
 spectral luminosity $L_\nu$\,, isotropic energy $E$ and $\rm DM_E$
 (extragalactic DM) mentioned in Sec.~\ref{sec1}.~As shown by the red
 lines in Fig.~\ref{fig1}, we find that these empirical relations still
 hold for 35 localized yFRBs, namely
 \bea
 &&\log E=0.8770\,\log L_\nu+10.2853\,,\quad {\rm with}\quad
 R=0.9695\,,\label{eq12}\\[1mm]
 &&\log L_\nu=2.7051\,\log {\rm DM_E}+26.9649\,,\quad {\rm with}\quad
 R=0.6756\,,\label{eq13}\\[1mm]
 &&\log E=2.6319\,\log {\rm DM_E}+33.3207\,,\quad {\rm with}
 \quad R=0.7266\,,\label{eq14}
 \eea
 with slopes and intercepts close to the ones in
 Eqs.~(\ref{eq3})\,--\,(\ref{eq5}).~The empirical relation~(\ref{eq12})
 is tight, while its $R$ value is higher than the one in
 Eq.~(4.8) of~\cite{Guo:2022wpf}.~The $R$ values for empirical
 relations~(\ref{eq13}) and (\ref{eq14}) are lower than the ones in
 Eqs.~(4.9) and (4.10) of~\cite{Guo:2022wpf}.~As shown by the blue lines
 in Fig.~\ref{fig1}, we also find similar empirical relations
 for 9 localized oFRBs, namely
 \bea
 &&\log E=0.7451\,\log L_\nu+14.8423\,,\quad {\rm with}\quad
 R=0.9251\,,\label{eq15}\\[1mm]
 &&\log L_\nu=1.7177\,\log {\rm DM_E}+29.5783\,,\quad {\rm with}\quad
 R=0.5414\,,\label{eq16}\\[1mm]
 &&\log E=1.1701\,\log {\rm DM_E}+37.1404\,,\quad {\rm with}
 \quad R=0.4579\,,\label{eq17}
 \eea
 but with quite different slopes and intercepts.~The empirical
 relations~(\ref{eq16}) and (\ref{eq17}) with low $R$ values
 are somewhat weak, mainly due to the number of localized
 oFRBs is too few.

In addition, it is of interest to find other empirical relations for
 the current localized FRBs.~We have tried many combinations of the
 observational and derived quantities.~In Fig.~\ref{fig2}, we present
 the other five empirical correlations with high score $R^2 > 0.7$ for
 both yFRBs and oFRBs.~As shown by the red lines in Fig.~\ref{fig2}, we
 find the following empirical relations for 35 localized yFRBs,
 \bea
 &&\log F_\nu=0.6875\,\log S_\nu+0.4410\,,\quad {\rm with}\quad
 R=0.9086\,,\label{eq18}\\[1mm]
 &&\log T_B=1.1697\,\log L_\nu-4.2276\,,\quad {\rm with}\quad
 R=0.9070\,,\label{eq19}\\[1mm]
 &&\log {\rm DM_E}=0.6459\,\log z+2.9285\,,\quad {\rm with}
 \quad R=0.8716\,,\label{eq20}\\[1mm]
 &&\log E=2.2648\,\log z+41.5195\,,\quad {\rm with}\quad
 R=0.8437\,,\label{eq21}\\[1mm]
 &&\log T_B=-3.4503\,\log W+35.6550\,,\quad {\rm with}\quad
 R=0.7936\,.\label{eq22}
 \eea
 There are similar empirical relations for 9 localized oFRBs as shown
 by the blue lines in Fig.~\ref{fig2}, but with quite different slopes
 and intercepts, namely
 \bea
 &&\log F_\nu=0.5673\,\log S_\nu+0.6790\,,\quad {\rm with}\quad
 R=0.8245\,,\label{eq23}\\[1mm]
 &&\log T_B=1.4759\,\log L_\nu-14.7987\,,\quad {\rm with}\quad
 R=0.9225\,,\label{eq24}\\[1mm]
 &&\log {\rm DM_E}=0.9582\,\log z+3.1489\,,\quad {\rm with}
 \quad R=0.7793\,,\label{eq25}\\[1mm]
 &&\log E=2.2743\,\log z+41.7648\,,\quad {\rm with}\quad
 R=0.7238\,,\label{eq26}\\[1mm]
 &&\log T_B=-3.5158\,\log W+36.1373\,,\quad {\rm with}\quad
 R=0.9222\,.\label{eq27}
 \eea
 The empirical relation~(\ref{eq22}) for 35 localized yFRBs, and the
 empirical relations~(\ref{eq25})\,--\,(\ref{eq26}) for 9 localized
 oFRBs, are somewhat weak as shown by their relatively low $R$ values.


 \begin{center}
 \begin{figure}[tb]
 \centering
 \vspace{-6mm}  
 \hspace{-6mm}  
 \includegraphics[width=0.98\textwidth]{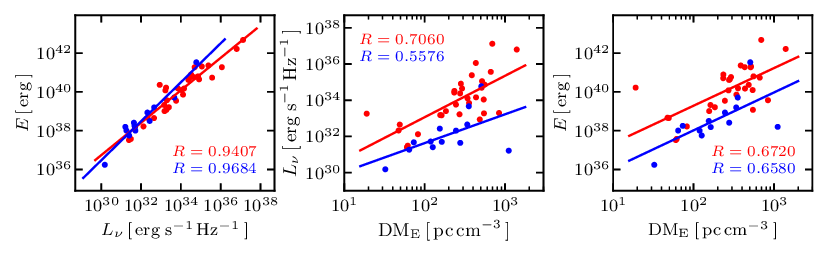}
 \vspace{-2mm}  
 \caption{\label{fig3} The empirical relations (red/blue
 lines) between spectral luminosity $L_\nu$, isotropic energy $E$ and
 $\rm DM_E$ for 31 localized nFRBs (red points) and 13 localized rFRBs
 (blue points), respectively.~The corresponding $R$ values (in
 red/blue) are also given. See Sec.~\ref{sec3b} for details.}
 \end{figure}
 \end{center}



 \begin{center}
 \begin{figure}[tb]
 \centering
 \vspace{-6mm}  
 \hspace{-6mm}  
 \includegraphics[width=0.98\textwidth]{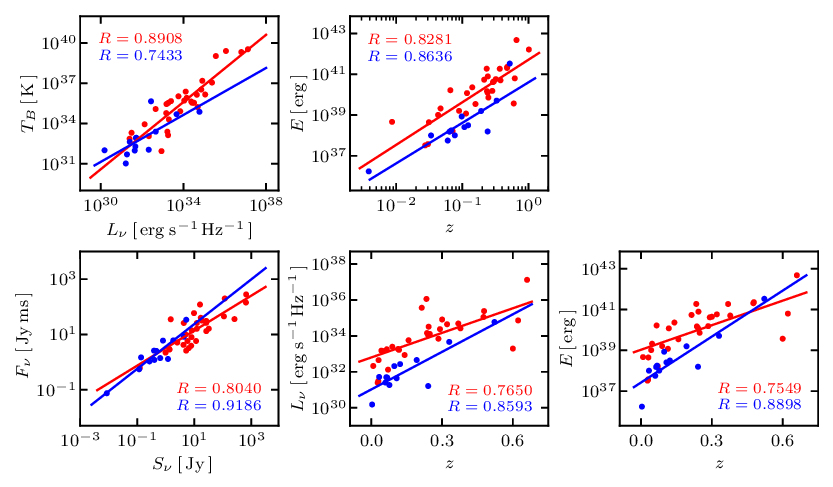}
 \vspace{-2mm}  
 \caption{\label{fig4} The other five empirical relations (red/blue
 lines) with high score $R^2 > 0.7$ for 31 localized nFRBs (red points)
 and 13 localized rFRBs (blue points), respectively.\hspace{1.28mm}The
 corresponding $R$ values (in red/blue) are also given. See
 Sec.~\ref{sec3b} for details.}
 \end{figure}
 \end{center}


\vspace{-19mm} 


\subsection{Empirical relations for the localized
 nFRBs and rFRBs}\label{sec3b}

Next, we also consider the empirical relations for 31 localized
 nFRBs (labeled with $\tt Rep.=0$) and 13 localized rFRBs (labeled with
 $\tt Rep.=1$) in Table~\ref{tab2}.

In Fig.~\ref{fig3}, we check the 2D empirical relations between
 spectral luminosity $L_\nu$\,, isotropic energy $E$ and $\rm DM_E$
 (extragalactic DM) mentioned in Sec.~\ref{sec1}.~As shown by the red
 lines in Fig.~\ref{fig3}, we find that these empirical relations still
 hold for 31 localized nFRBs, namely
 \bea
 &&\log E=0.8373\,\log L_\nu+11.6135\,,\quad {\rm with}\quad
 R=0.9407\,,\label{eq28}\\[1mm]
 &&\log L_\nu=2.2904\,\log {\rm DM_E}+28.4798\,,\quad {\rm with}\quad
 R=0.7060\,,\label{eq29}\\[1mm]
 &&\log E=1.9405\,\log {\rm DM_E}+35.4052\,,\quad {\rm with}
 \quad R=0.6720\,,\label{eq30}
 \eea
 with slopes and intercepts close to the ones in
 Eqs.~(\ref{eq3})\,--\,(\ref{eq5}).~There are similar empirical
 relations for 13 localized rFRBs as shown by the blue
 lines in Fig.~\ref{fig3}, namely
 \bea
 &&\log E=1.0066\,\log L_\nu+6.2684\,,\quad {\rm with}\quad
 R=0.9684\,,\label{eq31}\\[1mm]
 &&\log L_\nu=1.5926\,\log {\rm DM_E}+28.4448\,,\quad {\rm with}\quad
 R=0.5576\,,\label{eq32}\\[1mm]
 &&\log E=1.9537\,\log {\rm DM_E}+34.1080\,,\quad {\rm with}
 \quad R=0.6580\,,\label{eq33}
 \eea
 but with quite different slopes and intercepts.~The empirical
 $L_\nu-E$ relations for both nFRBs and rFRBs are tight, as shown by the
 high $R$ values.~But the empirical ${\rm DM_E}-L_\nu$ and
 ${\rm DM_E}-E$ relations for both nFRBs and rFRBs are somewhat weak,
 as shown by the relatively low $R$ values.

Again, we try to find more empirical relations for the current localized
 nFRBs and rFRBs.~In Fig.~\ref{fig4}, we present the other five
 empirical correlations with high score $R^2 > 0.7$ for both nFRBs and
 rFRBs.~As shown by the red lines in Fig.~\ref{fig4}, we find
 the following empirical relations for 31 localized nFRBs,
 \bea
 &&\log T_B=1.2536\,\log L_\nu-7.0191\,,\quad {\rm with}\quad
 R=0.8908\,,\label{eq34}\\[1mm]
 &&\log E=2.0861\,\log z+41.7115\,,\quad {\rm with}\quad
 R=0.8281\,,\label{eq35}\\[1mm]
 &&\log F_\nu=0.6328\,\log S_\nu+0.5049\,,\quad {\rm with}\quad
 R=0.8040\,,\label{eq36}\\[1mm]
 &&\log L_\nu=4.5683\,z+32.7988\,,\quad {\rm with}
 \quad R=0.7650\,,\label{eq37}\\[1mm]
 &&\log E=4.0127\,z+39.0268\,,\quad {\rm with}\quad
 R=0.7549\,.\label{eq38}
 \eea
 Note that the empirical relations~(\ref{eq35}) and (\ref{eq38}) are
 different in fact.~There are similar empirical relations
 for 13 localized rFRBs as shown by
 the blue lines in Fig.~\ref{fig4}, namely
 \bea
 &&\log T_B=0.8757\,\log L_\nu+4.8644\,,\quad {\rm with}\quad
 R=0.7433\,,\label{eq39}\\[1mm]
 &&\log E=1.9866\,\log z+40.6032\,,\quad {\rm with}\quad
 R=0.8636\,,\label{eq40}\\[1mm]
 &&\log F_\nu=0.8093\,\log S_\nu+0.5650\,,\quad {\rm with}\quad
 R=0.9186\,,\label{eq41}\\[1mm]
 &&\log L_\nu=6.9769\,z+31.0184\,,\quad {\rm with}
 \quad R=0.8593\,,\label{eq42}\\[1mm]
 &&\log E=7.5096\,z+37.4202\,,\quad {\rm with}\quad
 R=0.8898\,,\label{eq43}
 \eea
 but with quite different slopes and intercepts.~Please be aware of the
 difference between the empirical relations~(\ref{eq40})
 and (\ref{eq43}).


 \begin{center}
 \begin{figure}[tb]
 \centering
 \vspace{-6mm}  
 \hspace{-6mm}  
 \includegraphics[width=0.98\textwidth]{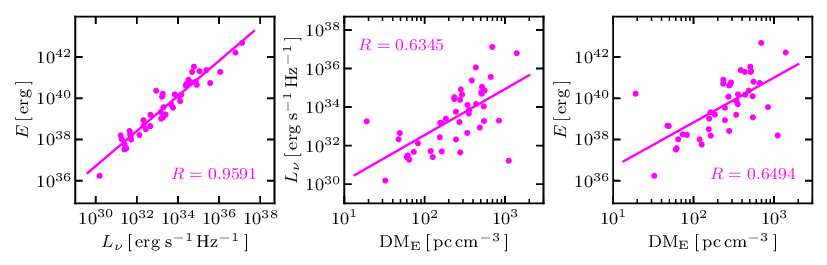}
 \vspace{-2mm}  
 \caption{\label{fig5} The empirical relations (magenta lines) between
 spectral luminosity $L_\nu$, isotropic energy $E$ and $\rm DM_E$ for
 all 44 localized FRBs (magenta points). The corresponding $R$
 values are also given. See Sec.~\ref{sec3c} for details.}
 \end{figure}
 \end{center}


\vspace{-14mm} 


\subsection{Empirical relations for
 all localized FRBs}\label{sec3c}

It is worth noting that the empirical $L_\nu-E$ relation holds
 solidly in all the cases of yFRBs, oFRBs, nFRBs, rFRBs considered in
 Secs.~\ref{sec3a} and \ref{sec3b}.~Actually, it is tight with high
 $R$ values close to 1, as shown in Eqs.~(\ref{eq12}), (\ref{eq15}),
 (\ref{eq28}) and (\ref{eq31}).~Noting that the empirical $L_\nu-E$
 relation is the key to calibrate FRBs as standard candles for
 cosmology, as mentioned in Sec.~\ref{sec1} (see~\cite{Guo:2023hgb} for
 details), it is of interest to check the empirical relations between
 spectral luminosity $L_\nu$\,, isotropic energy $E$ and $\rm DM_E$
 for all of the current 44 localized FRBs.~We
 present the results in Fig.~\ref{fig5}, and they are given by
 \bea
 &&\log E=0.8496\,\log L_\nu+11.2279\,,\quad {\rm with}\quad
 R=0.9591\,,\label{eq44}\\[1mm]
 &&\log L_\nu=2.3922\,\log {\rm DM_E}+27.7618\,,\quad {\rm with}\quad
 R=0.6345\,,\label{eq45}\\[1mm]
 &&\log E=2.1687\,\log {\rm DM_E}+34.4905\,,\quad {\rm with}
 \quad R=0.6494\,,\label{eq46}
 \eea
 with slopes and intercepts close to the ones in both the cases
 of yFRBs (n.b.~Eqs.~(\ref{eq12})\,--\,(\ref{eq14})) and nFRBs
 (n.b.~Eqs.~(\ref{eq28})\,--\,(\ref{eq30})).~Although the empirical
 ${\rm DM_E}-L_\nu$ and ${\rm DM_E}-E$ relations are somewhat weak with
 low $R$ values, the empirical $L_\nu-E$ relation stands firm with high
 $R$ value close to 1.


 \begin{center}
 \begin{figure}[tb]
 \centering
 \vspace{-6mm}  
 \hspace{-6mm}  
 \includegraphics[width=0.98\textwidth]{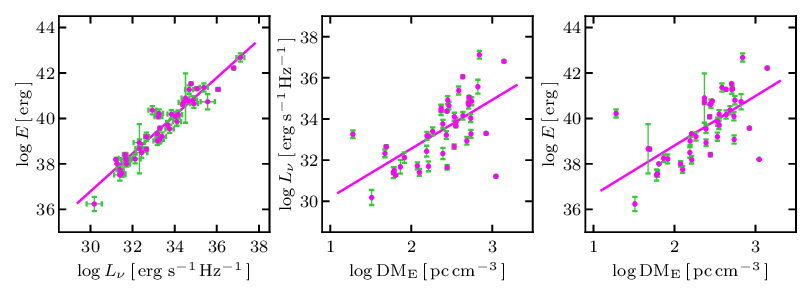}
 \vspace{-2mm}  
 \caption{\label{fig6} The linear empirical relations (magenta lines)
 between $\log L_\nu$\,, $\log E$ and $\log {\rm DM_E}$ for all
 of the 44 localized FRBs (magenta points) with both the horizontal and
 vertical error bars (in lime green). See Sec.~\ref{sec4} for details.}
 \end{figure}
 \end{center}



 \begin{center}
 \begin{figure}[tb]
 \centering
 \vspace{-0.3mm} 
 \hspace{-6mm}  
 \includegraphics[width=0.98\textwidth]{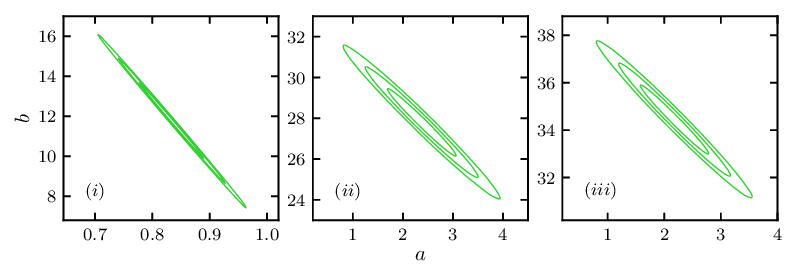}
 \vspace{-4.1mm} 
 \caption{\label{fig7} The $1\sigma$, $2\sigma$ and $3\sigma$ contours
 for slopes $a$ and intercepts $b$ of the linear empirical relations
 $(i)$~$\log L_\nu-\log E$, $(ii)$~$\log {\rm DM_E}-\log L_\nu$ and
 $(iii)$~$\log {\rm DM_E-\log E}$. See Sec.~\ref{sec4} for details.}
 \end{figure}
 \end{center}


\vspace{-19mm} 


\section{Taking uncertainties into account}\label{sec4}

In the previous section, the linear empirical relations
 are fitted to the data points without error bars by using
 {\tt sklearn.linear$_{-}$model.LinearRegression}
 in Python~\cite{LinearRegression}, which employs the ordinary least
 squares linear regression. In this section, we further take
 uncertainties into account.

Here, we do not try to consider all the empirical relations
 with uncertainties.~We mainly focus on the empirical relations between
 spectral luminosity $L_\nu$\,, isotropic energy $E$ and $\rm DM_E$,
 which are the main empirical relations found in~\cite{Guo:2022wpf}
 and the key to calibrate FRBs as standard candles for
 cosmology~\cite{Guo:2023hgb}. But we note that the method used here
 is very general and it can be easily apply to all empirical relations
 with uncertainties.

It is worth noting that $L_\nu$\,, $E$ and $\rm DM_E$ are all derived
 quantities.~When one takes their uncertainties into account, the error
 propagation should be considered carefully (see
 e.g.~\cite{Alam:2004ip,Wei:2006va,Liu:2014vda,Yin:2018mvu,
 Qiang:2020vta,errorp}).~Alternatively, one could use the Python package
 {\tt uncertainties}~\cite{uncertainties} to this end.~The observed
 quantities fluence $F_\nu$, flux $S_\nu$ and $\rm DM_{obs}$ are involved in
 $L_\nu$\,, $E$ and $\rm DM_E$ (n.b.~Eqs.~(\ref{eq1}) and (\ref{eq2})),
 while the uncertainties of redshifts $z$ and then $d_L$ are
 negligible.~Note that the uncertainties of $F_\nu$, $S_\nu$
 and $\rm DM_{obs}$ are given in Table~\ref{tab2}.~To be conservative,
 if the upper and lower errors are not equal, we adopt the
 larger one.~On the other hand, if the errors of some observed
 quantities are absent in Table~\ref{tab2}, we assign the
 average relative errors in the same columns to them.~Noting that the fluxes
 are absent for some FRBs in Table~\ref{tab2} and they can be estimated
 by using $\rm flux\simeq fluence/width$ as mentioned in
 Sec.~\ref{sec2}, the error propagation should also be
 considered here.~Finally, we obtain the uncertainties of
 $\log L_\nu$\,, $\log E$ and $\log {\rm DM_E}$ for all of the current
 44 localized FRBs, as shown by the horizontal and vertical error bars
 in Fig.~\ref{fig6}.~Note that the horizontal error bars of
 $\log {\rm DM_E}$ are too short to be seen by eyes, mainly due to the
 uncertainties of $\rm DM_{obs}$ for all FRBs are very small, as shown
 in Table~\ref{tab2}.

Next, the linear empirical relations will be fitted to these data points
 with both the horizontal and vertical error bars.~To our best
 knowledge, there are two main methods to this end in the
 literature.~One might use the bisector of the two ordinary
 least squares~\cite{Isobe:1990up} (see also e.g.~\cite{Schaefer:2006pa,
 Liang:2008kx,Wei:2008kq,Wei:2010wu,Liu:2014vda}).~Alternatively, in
 this work we use the Nukers' estimate~\cite{Tremaine:2002js} (see also
 e.g.~\cite{Tsvetkova:2017lea,Minaev:2019unh}).~If one fits the linear
 empirical relation with constant slope $a$ and intercept $b$, namely
 \be{eq47}
 y=ax+b\,,
 \ee
 to $N$ data points $(x_i,\ y_i)$ with errors $\sigma_{xi}$ and
 $\sigma_{yi}$, the Nukers' estimate is based on minimizing
 \be{eq48}
 \chi^2=\sum_{i\,=\,1}^N\,\frac{(y_i-ax_i-b)^2}
 {a^2\sigma_{xi}^2+\sigma_{yi}^2+\sigma_{\rm int}^2}\,,
 \ee
 where $\sigma_{\rm int}$ represents the intrinsic dispersion/scatter, which
 is determined by requiring $\chi^2/dof=1$~\cite{Tremaine:2002js} (see
 also e.g.~\cite{Tsvetkova:2017lea,Minaev:2019unh}).~One can
 minimize $\chi^2$ by using the Markov Chain Monte Carlo (MCMC) Python
 package {\tt emcee}~\cite{Foreman-Mackey:2012any} with
 {\tt GetDist}~\cite{Lewis:2019xzd} and then obtain the constraints
 on the free parameters $a$ and $b$.

By using this method, we fit the linear empirical relations between
 $\log L_\nu$\,, $\log E$ and $\log {\rm DM_E}$ to the 44 data points
 of all localized FRBs with both the horizontal and vertical error bars
 as shown in Fig.~\ref{fig6}. The results are given by
 \bea
 &&\log E=a\log L_\nu+b\quad {\rm with}\quad
 a=0.8336\pm 0.0375,\ b=11.7858\pm 1.2568,
 \ \sigma_{\rm int}=0.3308\,,\label{eq49}\\[1mm]
 &&\log L_\nu=a\log {\rm DM_E}+b\quad {\rm with}
 \quad a=2.3741\pm 0.4527,\ b=27.8054\pm 1.0893,
 \ \sigma_{\rm int}=1.2292\,,\hspace{10mm}\label{eq50}\\[1mm]
 &&\log E=a\log {\rm DM_E}+b\quad {\rm with}
 \quad a=2.1807\pm 0.3967,\ b=34.4443\pm 0.9568,
 \ \sigma_{\rm int}=1.0591\,,\label{eq51}
 \eea
 where the constraints on $a$ and $b$ are given by their means with
 $1\sigma$ uncertainties.~These linear empirical relations with their
 mean $a$ and $b$ are also plotted in Fig.~\ref{fig6}.~On the other
 hand, in Fig.~\ref{fig7}, we present the $1\sigma$, $2\sigma$
 and $3\sigma$ contours for $a$ and $b$ of these linear
 empirical relations.~From Eqs.~(\ref{eq49})\,--\,(\ref{eq51})
 and Fig.~\ref{fig7}, it is easy to see that the constraints on $a$ and
 $b$ are fairly tight.~Note that for the empirical $\log L_\nu-\log E$
 relation, $a=1$ is certainly far beyond $3\sigma$ region, and it is
 actually on the edge of $4\sigma$ region.~So, this is beneficial for
 using the empirical $\log L_\nu-\log E$ relation with $a\not=1$ to
 calibrate FRBs as standard candles for cosmology, as mentioned
 in Sec.~\ref{sec1} (see~\cite{Guo:2023hgb} for details).


 \begin{center}
 \begin{figure}[tb]
 \centering
 \vspace{-6mm} 
 \hspace{-6mm} 
 \includegraphics[width=0.72\textwidth]{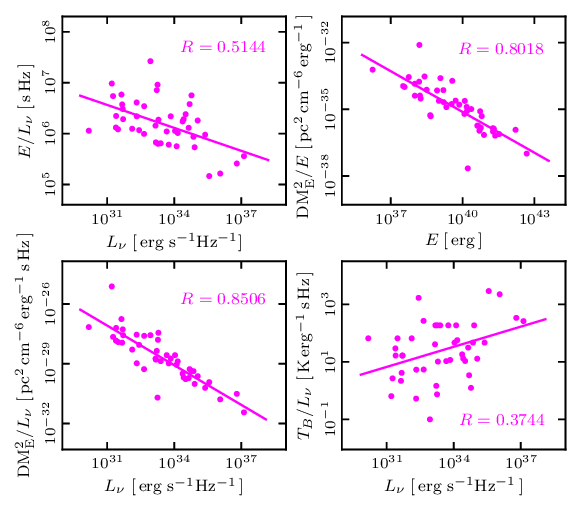}
 \vspace{-2mm} 
 \caption{\label{figindep2d} The empirical relations (magenta lines)
 between various independent quantities for all 44 localized
 FRBs (magenta points). The corresponding $R$ values are also
 given. See Sec.~\ref{secins} for details.}
 \end{figure}
 \end{center}



 \begin{center}
 \begin{figure}[tb]
 \centering
 \vspace{-12mm}  
 \hspace{-6mm}  
 \includegraphics[width=0.41\textwidth]{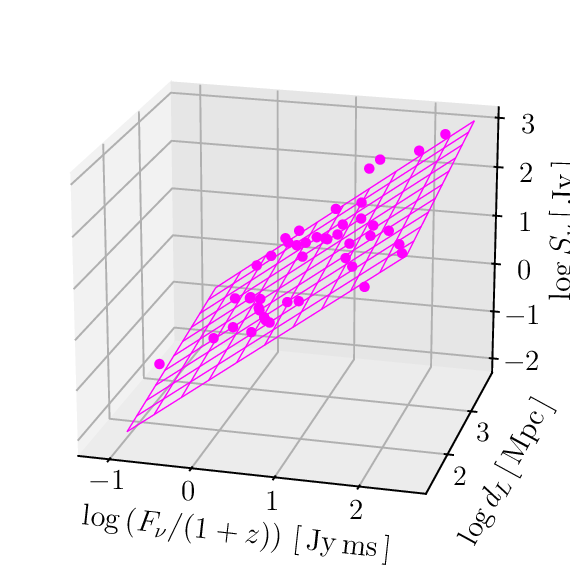}
 \vspace{-2mm} 
 \caption{\label{figindep3d} The 3D empirical relation (magenta meshed
 plane) between $S_\nu$, $d_L$ and $F_\nu/(1+z)$ for all 44 localized
 FRBs (magenta points). See Sec.~\ref{secins} for details.}
 \end{figure}
 \end{center}


\vspace{-19mm} 


\section{The independencies of
 empirical relations}\label{secins}

One might concern whether the empirical relations mentioned above are
 trivial or not.~The non-trivial relations involve independent
 quantities and usually carry physical meanings, as in the fields of
 e.g.~SNIa and GRBs (we thank the referee for pointing out this issue).
 In the above empirical relations, for example, $E$ and $L_\nu$ (or
 $F_\nu$ and $S_\nu$) are related to duration.~If the FRB
 duration distribution is narrow, a positive correlation between them
 can be expected.~On the other hand, the relation between $E$
 (or $L_\nu$) and $\rm DM_E$ (a proxy of luminosity distance $d_L$) is
 expected if $F_\nu$ (or $S_\nu$) has a narrow distribution.~Also, since
 $T_B$ is defined with $L_\nu$ and $W$, there are correlations between
 $T_B$ and $L_\nu$ or $W$ as expected.~Of course, the dependent indices
 are not simply the ones expected by the trivial relationships, as shown
 in the present work, but the physical origin of the deviation
 is unknown, and it could be introduced by some selection effects (we
 thank the referee for pointing out this issue).

These correlations should be carefully tested after the trivial dependencies
 are removed.~For example, instead of showing the relation between $E$
 and $L_\nu$, the relation between $E/L_\nu$ versus $L_\nu$ should be
 checked.~Similarly, the $T_B-L_\nu$ relation should be replaced by the
 $T_B/L_\nu$ versus $L_\nu$ relation.~In the cases of $\rm DM_E$ versus
 $E$ or $L_\nu$ relations, the $d_L$ dependencies also should
 be approximately removed (we thank the referee for this suggestion).~At
 first, we are interested in the empirical relations between
 $E$, $L_\nu$ and $\rm DM_E$, since they are important to calibrate FRBs
 as standard candles for cosmology, as mentioned above.~In
 Fig.~\ref{figindep2d}, we show the $E/L_\nu$ versus $L_\nu$,
 ${\rm DM_E^2}/E$ versus $E$, ${\rm DM_E^2}/L_\nu$ versus
 $L_\nu$ relations for all 44 localized FRBs.~Then, we also present the
 $T_B/L_\nu$ versus $L_\nu$ relation in Fig.~\ref{figindep2d}.~Noting
 that $E\propto d_L^2$, $L_\nu\propto d_L^2$ and $T_B\propto d_L^2$ by
 definitions in Eqs.~(\ref{eq1}) and (\ref{eq11}) respectively, while
 ${\rm DM_E}\sim d_L$ is a proxy of luminosity distance $d_L$, it is easy to
 see that $E/L_\nu$, ${\rm DM_E^2}/E$, ${\rm DM_E^2}/L_\nu$, $T_B/L_\nu$ are
 all independent of $d_L$ (at least approximately).~The
 numerical results are given by
 \bea
 &&\log\left(E/L_\nu\right)=-0.1504\,\log L_\nu+11.2279\,,
 \quad {\rm with}\quad R=0.5144\,,\label{eqindep2d1}\\[1mm]
 &&\log\left({\rm DM_E^2}/E\right)=-0.6111\,\log E-10.6759\,,
 \quad {\rm with}\quad R=0.8018\,,\label{eqindep2d2}\\[1mm]
 &&\log\left({\rm DM_E^2}/L_\nu\right)=-0.6634\,\log L_\nu-6.5167\,,
 \quad {\rm with}\quad R=0.8506\,,\label{eqindep2d3}\\[1mm]
 &&\log\left(T_B/L_\nu\right)=0.2336\,\log L_\nu-6.4169\,,\quad
 {\rm with}\quad R=0.3744\,.\label{eqindep2d4}
 \eea
 Obviously, the $E/L_\nu$ versus $L_\nu$, $T_B/L_\nu$ versus $L_\nu$
 relations become weaker, since their $R$ values are low, in comparison
 with the $R$ values for them in Fig.~\ref{fig5}, Figs.~\ref{fig2} and
 \ref{fig4}, respectively.~On the contrary, the ${\rm DM_E^2}/E$ versus
 $E$, ${\rm DM_E^2}/L_\nu$ versus $L_\nu$ relations become stronger,
 because their $R$ values are high, in comparison with the ones
 in Fig.~\ref{fig5}.~We stress again that $E/L_\nu$, ${\rm DM_E^2}/E$,
 ${\rm DM_E^2}/L_\nu$, $T_B/L_\nu$ are all independent of $d_L$ (at
 least approximately), as mentioned above.~So, the situation is
 still complicated and unclear.

Let us look it closely.~For instance, we can easily see that $d_L^2$ in both
 $E$ and $L_\nu$ (n.b.~Eq.~(\ref{eq1})) has been canceled in $E/L_\nu$,
 and hence it does not appear in Eq.~(\ref{eqindep2d1}), while the $R$ value
 for Eq.~(\ref{eqindep2d1}) is low.~Thus, it is reasonable to wonder whether
 the real correlation behind the $E$ versus $L_\nu$ relation instead lies in
 $d_L$ and other independent quantities.~In fact, the $E$ versus $L_\nu$
 relation in Eq.~(\ref{eq9}) turns out to be a 3D relation,
 namely $\log S_\nu=A\hspace{0.2mm}\log\left(F_\nu/(1+z)\right)
 +B\hspace{0.24mm}\log d_L+C$.~Fitting it to all 44 localized FRBs,
 we find
 \be{eqindep3d1}
 \log S_\nu=1.1927\,\log\left(F_\nu/(1+z)\right)+
 0.2849\,\log d_L-1.1555\,,\quad {\rm with}\quad R=0.8859\,,
 \ee
 and we also present it in Fig.~\ref{figindep3d}.~Clearly, the $R$ value for
 this 3D relation is fairly high.~By definition, the fluence $F_\nu$ is
 an integral of the flux $S_\nu$ with respect to the time $t$, and hence
 they are independent in principle.~Only in the case of short duration,
 $F_\nu\simeq S_\nu W$, and hence $F_\nu\propto S_\nu$, namely
 the coefficients in the $\log F_\nu-\log S_\nu$ relation should be 1.
 However, from Eqs.~(\ref{eq18}), (\ref{eq23}), (\ref{eq36})
 and (\ref{eq41}), one can easily see that it is not the case. In this work,
 we find that $F_\nu\propto S_\nu^\gamma$ and $\gamma\not=1$, namely it
 is rather a power-law relation.~Notice that in~\cite{Li:2021yds}, Li
 {\it et al.}~also found a tight power-law empirical relation between
 $F_\nu$ and $S_\nu$. So, it is better to regard $F_\nu$ and $S_\nu$ as
 independent quantities, and hence Eq.~(\ref{eqindep3d1}) is a tight 3D
 relation between independent quantities. Even if one does not
 agree this point and insists on $F_\nu\simeq S_\nu W$, we can instead
 consider the 3D relation $\log\left(W/(1+z)\right)=A\hspace{0.2mm}\log
 d_L+B\hspace{0.24mm}\log S_\nu+C$.~In this case, $W$, $d_L$, $S_\nu$
 are completely independent as well known.~Fitting it to
 all 44 localized FRBs, we find
 \be{eqindep3d2}
 \log\left(W/(1+z)\right)=-0.2099\,\log d_L-0.3354\,\log S_\nu+1.0030\,,
 \quad {\rm with}\quad R=0.6839\,.
 \ee
 It is still tight enough.~Note that if one replaces $d_L$ with
 $\rm DM_E$ in Eqs.~(\ref{eqindep3d1}) and (\ref{eqindep3d2}), they
 still work well, since $\rm DM_E$ is a proxy of luminosity distance
 $d_L$.~Now, let us try to understand the ``\,counter-intuitive\,''
 results in Eqs.~(\ref{eqindep2d2}) and (\ref{eqindep2d3}).~Although
 ${\rm DM_E^2}/E$ and ${\rm DM_E^2}/L_\nu$ are (approximately)
 independent of $d_L$ as mentioned above, noting that $d_L$,
 $F_\nu/(1+z)$, $S_\nu$ are involved in $E$ and $L_\nu$ by definitions
 (n.b.~Eq.~(\ref{eq1})), we find that Eqs.~(\ref{eqindep2d2})
 and (\ref{eqindep2d3}) turn out to be something
 like Eq.~(\ref{eqindep3d1}), while $\rm DM_E$ is not exactly $d_L$.~So,
 they boil down to 3D relations between $d_L$, $S_\nu$ and $F_\nu/(1+z)$
 or $W/(1+z)$ in Eqs.~(\ref{eqindep3d1}) or (\ref{eqindep3d2}), which
 are non-trivial as mentioned above.

In summary, there are at least several 3D empirical relations between
 independent quantities.~They are non-trivial.~Of course, a
 large uniform sample of localized FRBs from a single telescope/array is
 needed in the future to exclude selection effects.~To date, we do not
 know the physical mechanisms for these non-trivial empirical relations
 between independent quantities.~But let us keep an open mind.


\section{Concluding remarks}\label{sec5}

Although FRBs were discovered more than a decade ago, and they have been
 one of the active fields in astronomy and cosmology, their origins are
 still unknown.~An interesting topic closely related to the origins of FRBs
 is their classifications.~Different classes of FRBs require different
 physical mechanisms. If some empirical relations are found for
 different classes of FRBs, they might justify the classifications scenario
 and help us to reveal the physical mechanisms behind.~On the other
 hand, FRBs are actually a promising probe for cosmology, since their
 redshifts could be $z\sim 3$ or even higher.~Similar to the cosmology
 of SNIa or GRBs, some empirical relations might also play an important
 role in the FRB cosmology. In the literature, some new classifications
 of FRBs different from repeaters and non-repeaters were proposed recently.
 In particular, it was suggested to classify FRBs into the ones
 associated with old or young stellar populations, and some empirical
 relations have also been found for them, respectively. One of these
 empirical relations (namely $L_\nu-E$ relation) without DM has been
 used to calibrate FRBs as standard candles for cosmology.~This shows
 the potential of the new classification and the empirical relations for
 FRBs.~Nowadays, more than 50 FRBs have been well localized, and hence
 their redshifts $z$ are observationally known.~So, it is of interest to
 check the empirical relations with the actual data of current localized
 FRBs.~We find that many empirical relations still hold, and in
 particular the one used to calibrate FRBs as standard candles
 for cosmology stands firm.~This is beneficial to the FRB cosmology.

Some remarks are in order.~Actually, the number of current $44\sim 52$
 localized FRBs is not enough.~If the number of well-localized FRBs can
 be more than ${\cal O}(10^2\sim 10^3)$ in the near future, and if a
 large uniform sample of localized FRBs from a single telescope/array is
 available at that time, we might eliminate the possibility that the
 selection effects lead to the ``\,artificial\,'' empirical relations.~Let
 us keep an open mind and wait with patience.~But this does not deny
 the preliminary results in the present work, which shed promising light
 on the empirical relations for FRBs, and also give us a hopeful future.

If these empirical relations are real, they are meaningful on
 two sides.~First, they might support the other classifications of FRBs
 different from repeaters and non-repeaters, especially the new
 classification oFRBs/yFRBs proposed in~\cite{Guo:2022wpf,
 Guo:2023hgb}.~Second, they might help FRBs to be a promising probe for
 cosmology. In particular, the empirical $L_\nu-E$ relation stands firm
 with the current localized FRBs, and the slope $a\not=1$ in
 Eq.~(\ref{eq9}) far beyond $3\sigma$ confidence level (C.L.), actually
 on the edge of $4\sigma$ C.L., as shown in Sec.~\ref{sec4} of this
 work. If $a=1$, the luminosity distance $d_L$ will be canceled in both
 sides of Eq.~(\ref{eq9}) (n.b.~Eq.~(\ref{eq1})), and hence it cannot be
 used to study cosmology.~But actually $a\not=1$ as shown in this work
 by using the current localized FRBs, and then Eq.~(\ref{eq9}) can be
 recast as~\cite{Guo:2023hgb}
 \vspace{-0.1mm} 
 \be{eq52}
 \mu=-\frac{5}{2\left(1-\alpha\right)}\,\log\frac{\;F_\nu/(1+z)\,}{\rm
 Jy\,ms}+\frac{5\alpha}{2\left(1-\alpha\right)}\,\log\frac{S_\nu}{\;\rm
 Jy\,}+\beta\,,
 \ee
 where $\alpha=a\not=1$, $\beta=const.$ is a complicated combination of
 $a$ and $b$, while $\mu$ is the well-known distance modulus defined by
 \be{eq53}
 \mu=5\log\frac{d_L}{\,\rm Mpc}+25\,.
 \ee
 As shown in~\cite{Guo:2023hgb} by simulations, Eq.~(\ref{eq52}) can be
 used to calibrate the localized FRBs as standard candles for cosmology,
 complementarily to e.g.~SNIa, CMB and GRBs.

Note that for nearby FRBs such as FRB 20200120E and FRB
 20220319D, $L_\nu$ and $E$ can be calculated by using the measured
 luminosity distances of the sources.~The correlations not involving
 $\rm DM_E$ can also be investigated (we thank the referee for pointing
 out this issue).~However, as mentioned in the beginning of Sec.~\ref{sec2},
 in this work we exclude FRB 20200120E for its redshift $z<0$, and FRB
 20220319D for its $\rm DM_E<0$.~Thus, we have not used them in studying
 the empirical relations.~But in the future, one might consider other
 new nearby FRBs to this end.

As is discussed in Sec.~\ref{secins}, one should be aware of the caveat
 that some empirical relations found in this work and~\cite{Guo:2022wpf,
 Guo:2023hgb} might not involve independent quantities, and hence they
 might be trivial actually.~On the other hand, we also stress that there
 might be several non-trivial (especially 3D) relations between
 independent quantities, as in Eqs.~(\ref{eqindep3d1}) and
 (\ref{eqindep3d2}).~So far, we cannot settle the debate.~A
 large uniform sample of localized FRBs from a single telescope/array is
 needed in the future to this end.~Let us keep an open mind.

Finally, different physical mechanisms are required by
 different classes of FRBs, and different classes of FRBs have different
 empirical relations.~Obviously, the empirical relations of FRBs require
 some physical mechanisms behind them.~Therefore, they might help us
 to reveal the origins of FRBs.~These topics are cross-related, and
 actually have important values in this field.


\section*{ACKNOWLEDGEMENTS}

We thank the anonymous referee for quite useful comments and
 suggestions, which helped us to improve this work.~We are grateful to
 Han-Yue~Guo, Jia-Lei~Niu, Yun-Long~Wang, Shu-Yan~Long, Hui-Qiang~Liu
 and Yu-Xuan~Li for kind help and useful discussions.~This work was
 supported in part by NSFC under Grants No.~12375042 and No.~11975046.

\renewcommand{\baselinestretch}{1.1}


\end{document}